\documentclass[11pt]{article}

\usepackage[utf8]{inputenc}
\usepackage[T1]{fontenc}
\usepackage{geometry}
\usepackage{graphicx}
\usepackage{subcaption}
\usepackage{booktabs}
\usepackage{multirow}
\usepackage{hyperref}
\usepackage{amsmath}
\usepackage{xcolor}
\usepackage{natbib}
\usepackage{authblk}
\usepackage{tikz}
\usepackage{placeins}
\usetikzlibrary{positioning, shapes.geometric, arrows.meta, fit, calc}

\newcommand\blfootnote[1]{\begingroup\renewcommand\thefootnote{}\footnotetext{#1}\endgroup}

\title{Dendrite: A Real-Time Python Application for Online Brain-Computer Interface Research and Development}

\author[1]{Niko Kroflic}
\author[1]{Jan Babič}
\affil[1]{Laboratory for Neuromechanics and Biorobotics, Department of Automatics, Biocybernetics, and Robotics, Jožef Stefan Institute, Ljubljana, Slovenia}

\date{}

\begin{document}

\maketitle

\blfootnote{Preprint. This manuscript has not yet undergone peer review}

\begin{abstract}
Online brain-computer interface research requires software that can acquire multimodal physiological data, train and update decoders, run live inference, and preserve the full experimental provenance in a reproducible workflow. We present Dendrite, a real-time brain-computer interface application in Python that brings signal acquisition, decoder training, and live inference together in a single, ready-to-run application that stays modifiable. Dendrite records several signal streams at once, each at its native rate, and executes multiple processing modes concurrently against them. A decoder can start from a previously trained model or be fit mid-session while the pipeline keeps running, and the same recordings feed offline training in the same application. Each recording, decoder, and training run is tracked in a database, and every decoder records the configuration and the source recordings it was trained from, so a deployed decoder traces back to what produced it. The experimental paradigm stays external, an independent program in any language that reaches Dendrite over the network, rather than a module built inside the runtime. We validate the full system end-to-end on in-house and public BCI datasets, training and updating decoders online while the pipeline runs in real time. Dendrite is open-source under the GPL-3.0 license at \url{https://github.com/dendrite-bci/dendrite}. The result is a reproducible, open-source biomedical-computing system for developing and evaluating online BCI paradigms.

\end{abstract}

\section{Introduction}
\label{sec:intro}

Brain-computer interface (BCI) research is iterative. A study usually designs a task paradigm (the calibration routine, stimulus protocol, and trial structure that drive a session), acquires neural data under it, trains a decoder offline, deploys that decoder for online inference, and revises the paradigm for the next session. Supporting this loop asks several things of the software at once. It must acquire multiple signal streams, train and update decoders from those recordings, decode the streams in real time, and let the experimental paradigm evolve independently of the runtime. Compounding this, requirements may shift from one study to the next. Preprocessing, decoder choices, and output protocols should therefore be adjustable, so the runtime is expected to be modifiable by the researcher.

From a biomedical-computing perspective, this fragmentation has practical consequences. Decoder performance, latency, preprocessing choices, data provenance, and task timing are often distributed across separate scripts, acquisition programs, and offline analysis environments. This makes it difficult to reproduce an online BCI session, compare configurations under identical runtime conditions, or trace a deployed decoder back to the exact data and parameters that produced it. A useful BCI software system should therefore run in real time and preserve the complete path from acquisition to training, deployment, replay, and post-hoc analysis.

Existing tools trade off how much of a working system they supply against how tightly the experimental paradigm is bound to it. At one end, full platforms such as BCI2000~\citep{schalk2004bci2000}, MEDUSA~\citep{santamaria2023medusa}, MetaBCI~\citep{mei2024metabci}, and OpenViBE~\citep{renard2010openvibe} ship acquisition, processing, and a real-time runtime, and are usable the day they are installed. Most accept external stimulus software, integrated with the platform as an application module, a paradigm class, or a scenario. At the other end, the scientific-Python stack that BCI research already relies on supplies methods rather than a system: MNE~\citep{gramfort2013mne} for preprocessing and analysis, scikit-learn~\citep{pedregosa2011sklearn} for classical pipelines, Braindecode~\citep{schirrmeister2017braindecode} for neural decoders, and MOABB~\citep{jayaram2018moabb} for reproducible benchmarking. These are offline-first and leave acquisition, the runtime, and the online loop for the researcher to assemble.

Dendrite is a real-time Python BCI application built around three design goals. It performs multi-modal acquisition and decoding at native rates, unifies online and offline analysis in one launchable application, and keeps analysis and experiment setup low-friction, with the experimental paradigm kept external behind a Lab Streaming Layer (LSL)~\citep{kothe2025lsl} event contract. Because the data science and machine learning libraries BCI research already uses are Python, implementing the runtime in Python keeps the whole pipeline modifiable in the language researchers already work in, and lets the same processing and decoder code run against a recorded replay or a live amplifier.

Internally the runtime divides into four layers, for data, processing, machine learning, and the web interface. Three processing modes run in the processing layer, each in its own subprocess and each reading one shared-memory ring buffer. The synchronous mode extracts event-locked epochs, the asynchronous mode classifies a continuous sliding window, and the neurofeedback mode estimates band power. Because these modes share state rather than a fixed pipeline graph, a decoder trained by one mode is available to another within the same session, with no break in output. Recordings are written to HDF5 with metadata inspired by the Brain Imaging Data Structure (BIDS)~\citep{gorgolewski2016bids} and indexed in SQLite, so a completed session can be loaded directly into the training workbench and trained into a decoder that can be returned to the live pipeline. Training, live monitoring, data browsing, and offline replay all run in the same browser-served single-page application.

Dendrite makes three contributions for reproducible online BCI research.

\begin{itemize}
\item \textbf{Per-stream multimodal acquisition and decoding.} Per-stream ring buffers record any LSL stream at its native rate, a mode binds to each stream, and decoders fuse at the prediction layer rather than through joint training.
\item \textbf{Online and offline analysis inside one application.} A decoder trained mid-session by one processing mode is swapped into a second, independently running mode between inference steps, so live prediction continues without interruption while the model is replaced, and the same recordings feed offline training and Optuna hyperparameter search in the same application, with no second tool to install.
\item \textbf{Low-friction analysis and experiment setup.} The experimental paradigm is any program in any language and reaches Dendrite only through an LSL-event contract, and the same pipeline configuration drives a live amplifier or a recorded replay. Each session is written as a self-describing HDF5 recording with BIDS-inspired identifiers and indexed in SQLite, so a completed session loads directly into the training workbench with no conversion step, exports to MNE and FIF for offline analysis in the standard Python stack, and every trained decoder records, in the database, the training run and configuration that produced it together with the identifiers of its source recordings. Configuration, decoder training, and live monitoring all run from any browser on the local network.
\end{itemize}

\section{System Architecture}
\label{sec:architecture}

\subsection{Overview}

\begin{figure}[h]
\centering
\includegraphics[width=\textwidth]{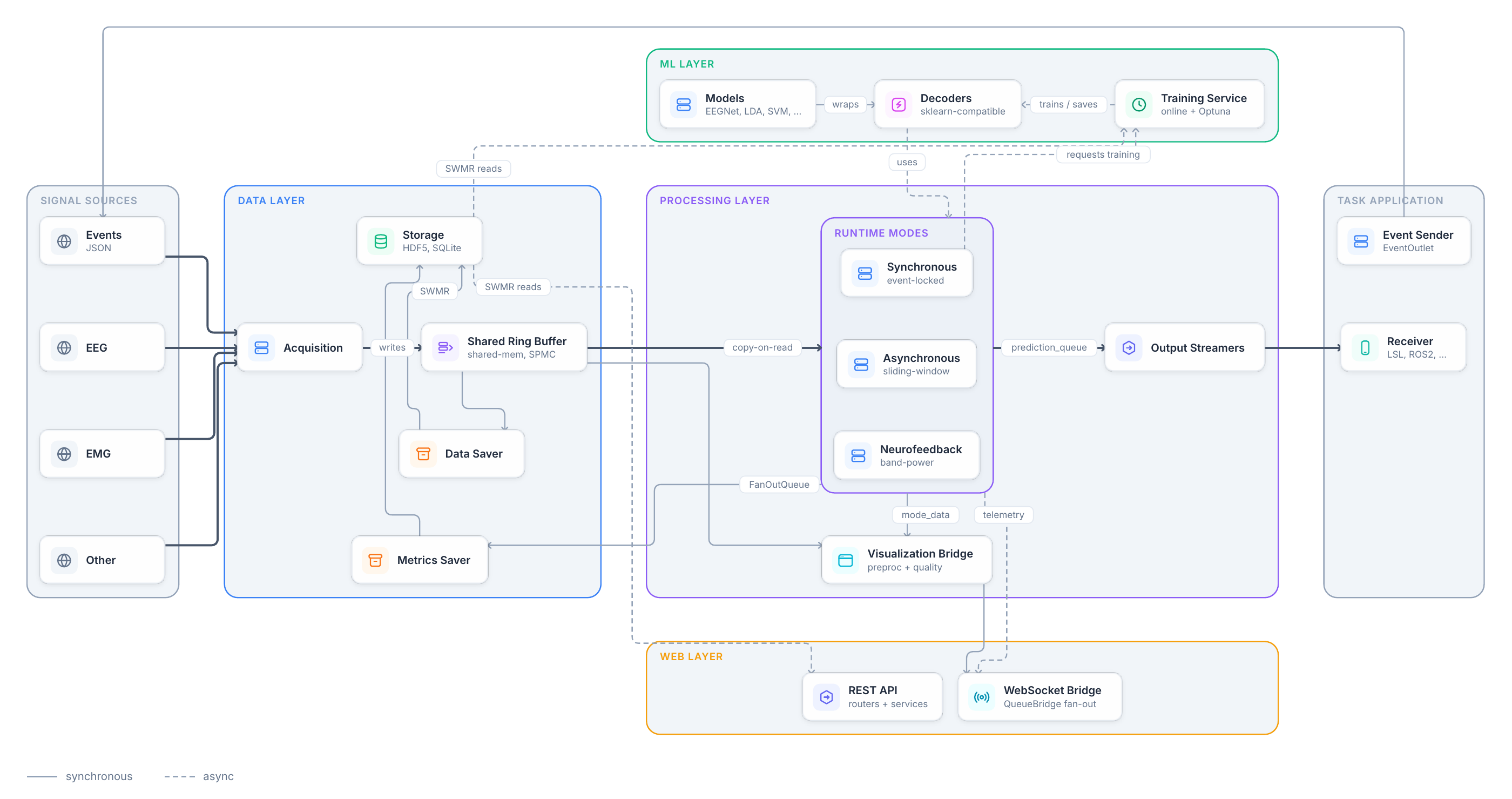}
\caption{Dendrite has four layers plus an LSL-event contract for external task applications. Multimodal inputs (EEG, EMG, events, kinematics) enter the Data Layer for acquisition and HDF5/SQLite storage, then reach the Processing Layer, where a shared ring buffer feeds composable runtime modes. The Machine Learning Layer wraps the decoder models, the sklearn-compatible decoders, and the online training service. The FastAPI Web Layer exposes a REST API and a WebSocket bridge (QueueBridge fan-out) that serve the Vue 3 single-page application. External applications receive predictions via configurable output protocols (LSL, ROS2, TCP/UDP, ZeroMQ).}
\label{fig:architecture}
\end{figure}

The architecture was designed around the complete biomedical BCI workflow, from synchronised acquisition of heterogeneous physiological streams through real-time processing, decoder training, and online deployment to reproducible replay from stored sessions. Dendrite is organised as four layers plus an LSL-event contract for external task applications (Figure~\ref{fig:architecture}). The Data Layer handles acquisition and storage, the Processing Layer runs the signal processing modes, the Machine Learning Layer provides the decoder infrastructure and training service, and the Web Interface Layer, served from the same backend process as the API, is the operator's surface. The Task Application boundary is not a layer but a contract, over which external paradigm software interacts with Dendrite through LSL events and a prediction stream. Each layer exposes only a narrow interface to its neighbours. Hardware-agnostic streams reach the Data Layer, samples reach the Processing Layer over a shared ring buffer, standardised decoder objects reach the Machine Learning Layer, REST and WebSocket channels reach the Web Interface, and LSL events and a prediction stream cross the Task Application boundary.

\subsection{Data layer}

The Data Layer ingests multimodal streams and synchronises signals at different sampling rates. Hardware-agnostic stream discovery and network transport come from LSL, so Dendrite connects to any LSL-compatible amplifier (g.tec, Brain Products, OpenBCI, Muse, custom hardware) without device-specific code. Stream metadata (channel labels, types, sampling rates) comes from LSL headers. For development without hardware, an offline replay manager (Section~\ref{sec:web}) plays recorded files back as live LSL streams, so the same code path runs against a saved session or a live amplifier.

The DAQ acquires each stream at its own native rate, from high-rate EEG and EMG down to irregularly timed event markers. Each stream runs in a dedicated acquisition thread that pulls samples at the native rate and writes them into a per-stream shared-memory ring buffer.

The ring buffer is a zero-copy structure in Python shared memory that one writer fills and many readers drain. Acquisition is the sole writer, while the data saver, the visualisation bridge, and every active processing mode read independently without inter-process queues. A slow consumer cannot stall acquisition or its peers. Each sample carries three timestamps (LSL-synchronised, local clock, wall-clock receive time) for latency tracking and offline alignment. LSL's clock-synchronisation protocol corrects timing offsets between devices, enabling direct temporal comparison across streams from different hardware. Event markers arrive on a separate LSL stream and are injected into a markers channel aligned with all data streams, which supports event-locked epoch extraction across modalities.

Raw data flows to HDF5 via a dedicated saver process. Each recording carries BIDS-inspired identifiers (subject, session, run, task)~\citep{gorgolewski2016bids} in its root attributes. The hierarchical layout stores per-modality datasets with structured arrays containing channel data and timestamps. After the first flush the saver enables HDF5 single-writer / multi-reader (SWMR) mode, so the ML Workbench, the Data Explorer, and external scripts can read up to the latest flushed sample while the recording is in progress, without copying or pausing acquisition. A SQLite backend tracks experiment lineage across four tables for studies, recordings, trained decoders, and training jobs. Each recording links to its acquisition configuration and any derived datasets or decoders. Trained decoders carry versioning metadata (accuracy, training parameters, data sources). This design makes each online session auditable. Raw signals, event markers, acquisition parameters, trained decoders, and performance metadata remain linked after the session, so an online decision made by the system can be traced back to the recording and configuration that generated the corresponding decoder.

\subsection{Processing layer and concurrency model}

The Processing Layer reads from the shared ring buffer and runs each processing mode in its own subprocess. Modes manage their own working buffers and publish predictions through a unified output packet. Process isolation prevents a crash in one mode from affecting the others. This matters in online BCI experiments, where acquisition should continue even if a decoder, visualisation, or experimental processing mode fails.

The three current modes address complementary paradigms. The synchronous mode targets event-triggered epochs in cued tasks, the asynchronous mode continuous decoding in self-paced control, and the neurofeedback mode spectral features in volitional modulation training. Any combination runs against the same source streams. A typical configuration pairs synchronous training with asynchronous inference so a newly fitted decoder hot-swaps into live inference without interrupting output.

Each mode owns its preprocessing chain rather than sharing a system-wide pipeline, so two modes can apply different bandpasses, references, or downsampling to the same source data. Modality processors maintain filter state across chunks for continuous real-time operation. EEG preprocessing supports configurable causal IIR bandpass and notch filtering, common-average re-referencing with bad-channel exclusion, and anti-aliased downsampling. Additional modality processors register through a common interface. Because all modes pull from the same shared ring buffer, sample distribution requires no per-mode queues and no fan-out logic on the producer side. Backpressure applies only at the visualisation output, where a slow browser client causes frames destined for the visualisation and mode-data channels to be dropped rather than buffered. The pipeline never blocks on the UI.

\subsubsection{Synchronous mode}

Synchronous mode handles event-locked decoding. It extracts a post-stimulus epoch on each marker, makes a per-trial prediction once a decoder is available, and triggers retraining as new epochs accumulate. When an event marker arrives, the mode schedules a window for extraction from its working buffer once enough post-stimulus data has arrived. The integer event code maps to a class label through the per-mode configuration. After every $N$ completed epochs, the mode sends a training request to the ML service. The service reads epoch data directly from the live recording file via SWMR concurrent access and runs training in a subprocess. Epoch ingestion continues uninterrupted. The new decoder is published into shared state, from which the synchronous mode reloads it for subsequent trials and any linked asynchronous mode picks it up for continuous inference. Each epoch yields a downsampled ERP for live visualisation, a per-trial prediction with confidence, and running performance metrics (accuracy, confidence, Cohen's kappa) for the operator dashboard.

\subsubsection{Asynchronous mode}

Asynchronous mode provides continuous classification using sliding windows. At configurable intervals it extracts a fixed-duration window and runs inference, emitting a per-window class prediction with confidence. Pre-trained models load at startup. When linked to a synchronous mode, the asynchronous mode picks up newly trained decoders from shared state and hot-swaps them without interrupting inference. External applications consume this prediction stream to drive closed-loop actions, applying whatever decision logic their paradigm requires.

\subsubsection{Neurofeedback mode}

Neurofeedback mode estimates spectral power inside one or more configurable frequency bands. The mode emits per-channel or cluster-averaged power values at a configurable update rate (4~Hz in the sessions reported here), which downstream applications convert into visual, auditory, or haptic feedback for attention training, relaxation, and motor rehabilitation protocols.

\subsection{Machine learning layer}
\label{sec:ml}

The Machine Learning Layer exposes a uniform decoder interface over two families of models, neural architectures (both native and adapted from Braindecode~\citep{schirrmeister2017braindecode}, such as EEGNet~\citep{lawhern2018eegnet}) and a classical decoder pipeline. Every decoder presents the same scikit-learn fit/predict interface, so cross-validation, MOABB~\citep{jayaram2018moabb} benchmarking, and Optuna~\citep{akiba2019optuna} hyperparameter search apply uniformly across the registry.

Neural and classical decoders share that interface but follow different training paths underneath. Neural pipelines delegate to a classifier that manages PyTorch~\citep{paszke2019pytorch} training loops, batch construction, and device placement across CPU, CUDA, and MPS backends. Classical pipelines call scikit-learn's fit directly. Both paths produce decoders that serialise to disk and load into real-time modes without modification. Decoders operate on single modalities. For multimodal experiments, users can run independent modes per signal type with predictions fused downstream. Dendrite does not require a specific decoder family and does not claim algorithmic novelty in the classifiers themselves. The contribution is the runtime integration, where the same decoder interface supports offline benchmarking, online retraining, live deployment, and storage of the trained model with the configuration and source recordings it was trained from.

\subsection{Task-application contract}
\label{sec:contract}

External applications (stimulus presentation software, game engines, custom scripts) handle paradigm logic such as stimulus timing and trial structure, and interact with Dendrite by creating an LSL outlet for event markers and subscribing to a prediction stream for closed-loop control. Each event is a JSON object carrying an integer event ID, a descriptive event type, and arbitrary contextual fields (trial, block, condition). The integer code triggers epoch extraction in synchronous mode and maps to a semantic class label through the mode configuration, and the full payload is persisted with the recording for offline analysis. Beyond discrete events, any numeric LSL stream at a fixed sampling rate can serve as a continuous input.

Modes publish predictions through a unified packet that carries the mode type, the prediction payload, and a timestamp. Task applications subscribe to a prediction stream and receive JSON-formatted results that drive stimulus presentation, game mechanics, or downstream control loops. Beyond LSL, Dendrite supports ROS2~\citep{macenski2022ros2} for robotics integration, TCP/UDP sockets, and ZeroMQ for distributed architectures. Multiple protocols can output simultaneously, so a single session can feed multiple consumers over different protocols at once from the same prediction stream.

\subsection{Web interface}
\label{sec:web}

A FastAPI backend serves both the REST API for configuration and the built Vue 3 single-page application, so the entire system is reachable from one LAN address with no client-side installation. Singleton services manage configuration, pipeline lifecycle, stream discovery, mode instances, ML jobs, database access, preflight validation, and offline replay-stream lifecycle, and they aggregate this state into a single pipeline configuration that the processing subprocess consumes at start.

\begin{figure}[h]
\centering
\includegraphics[width=\textwidth]{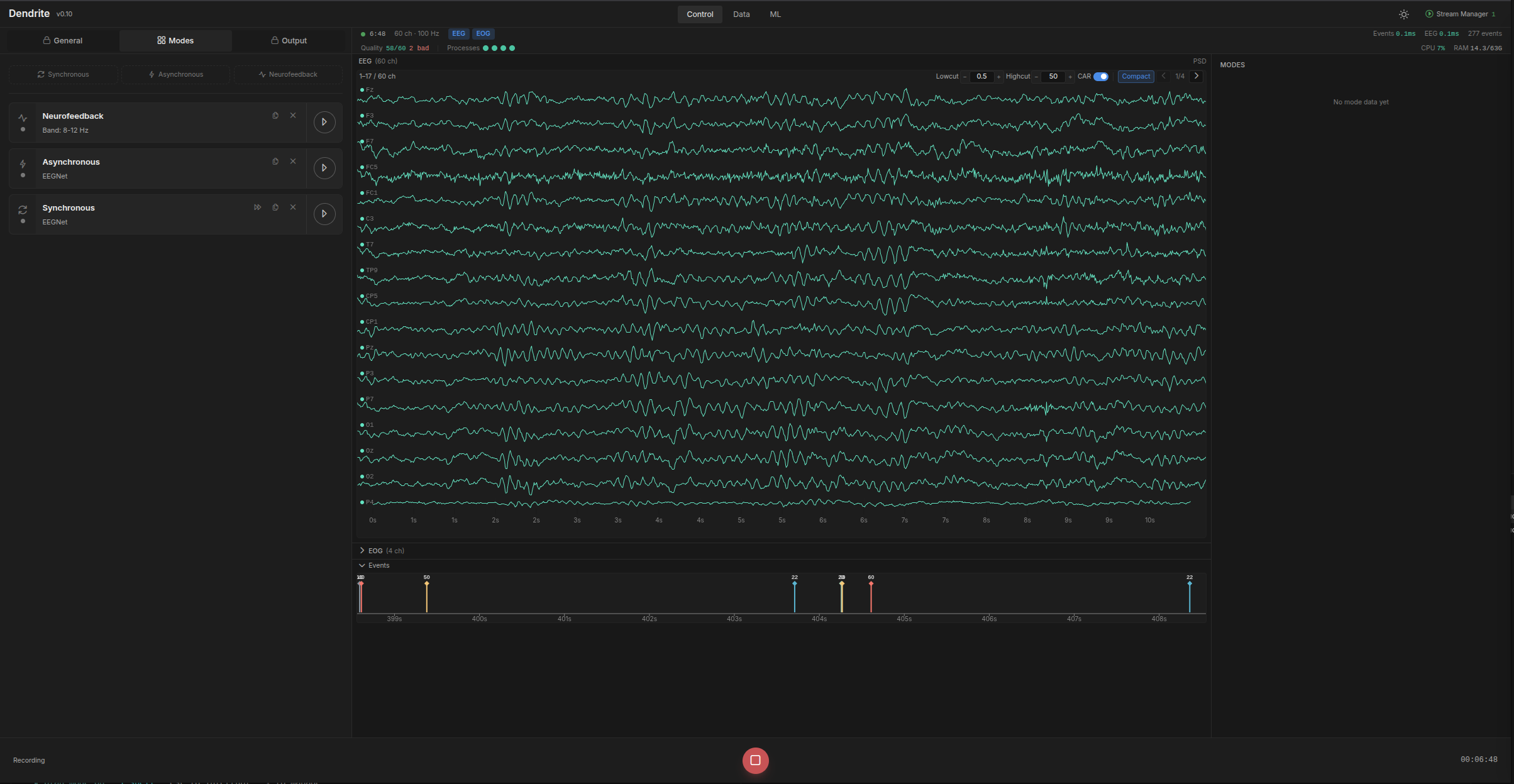}
\caption{Dendrite Control view (v0.10) during a live recording. The centre panel shows live multi-channel EEG, decimated and preprocessed for visualisation, with channel-quality colouring. The modes panel on the left lists three concurrently active modes (Neurofeedback for band power, Asynchronous for continuous inference, Synchronous for event-triggered training), all reading from the same shared ring buffer. The bottom strip shows the LSL event timeline. The view is served from the FastAPI backend over WebSocket channels and accessible from any browser on the LAN.}
\label{fig:dashboard}
\end{figure}

The frontend exposes three routed views (Figure~\ref{fig:dashboard}). \emph{Control} is the operator surface, a live dashboard for signal visualisation, channel quality, and mode outputs, alongside controls for stream selection and offline replay. \emph{Data Explorer} browses studies, recordings, and trained decoders persisted in SQLite. \emph{ML Workbench} drives offline training by loading recordings or MOABB datasets, configuring models, launching training jobs, and inspecting results. Real-time data reaches the browser over four WebSocket channels, namely telemetry (1~Hz JSON carrying system load, stream latency, mode metrics, and channel quality), visualisation ($\sim$100~Hz msgpack carrying the decimated preprocessed signal), mode-data (event-driven msgpack carrying mode predictions and band-power outputs), and training (per-epoch JSON carrying training progress and completion). Model training and Optuna trials run in dedicated subprocesses, so a long job never blocks the API or the UI.

\section{Evaluation}
\label{sec:evaluation}

\subsection{Setup}

The evaluation asks three software-system questions. Can Dendrite run online training and inference end-to-end? Can public and in-house datasets be replayed through the same runtime? Do inference latency and output cadence satisfy online BCI constraints? We exercise the full pipeline end-to-end with a reproducible harness that drives the live backend the way an operator would. It spawns the FastAPI server, configures one or more synchronous/asynchronous mode pairs (each sharing a decoder), and replays a recorded session over LSL through the offline replay manager. The synchronous mode retrains every ten epochs, each time fitting a fresh decoder on all epochs accumulated so far. The asynchronous mode hot-swaps each new decoder mid-session. We score the synchronous mode by \emph{prequential accuracy}, predicting every trial before its label is revealed and then folding it into the next retrain under a $0.95$ exponential forgetting factor. The reported final value is therefore a trailing weighted average rather than lifetime accuracy. The asynchronous mode is the runtime's online-deployment path, so we characterise it by the operational quantities a closed-loop application depends on, namely per-step inference latency and step cadence (Section~\ref{sec:latency}), rather than by a separate accuracy score. Because the asynchronous mode runs the same decoder the synchronous mode trains, its decoding quality tracks the synchronous prequential accuracy, and we treat that correspondence as a fidelity check rather than an independent result.

\begin{figure}[tbp]
\centering
\includegraphics[width=\textwidth]{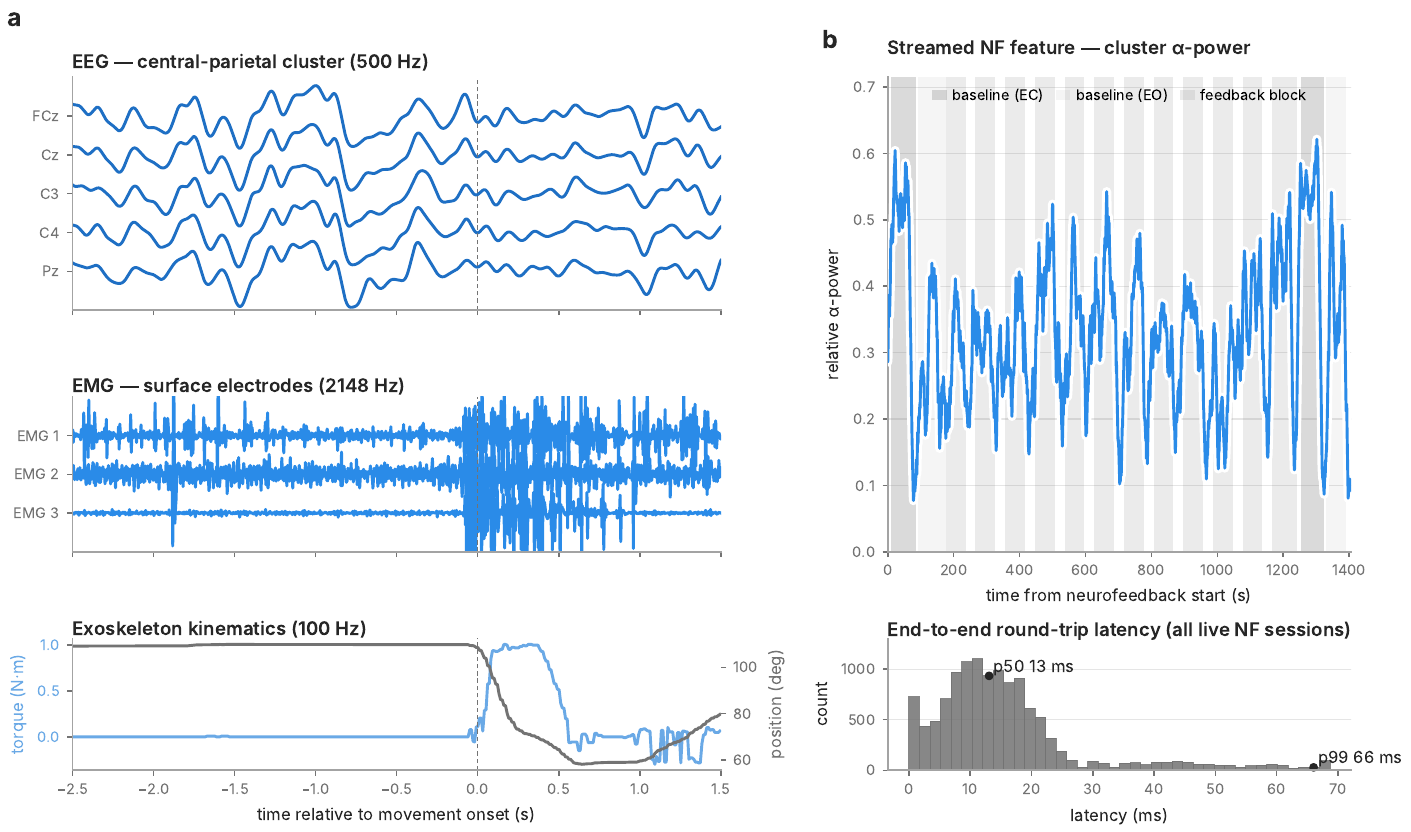}
\caption{Acquisition and closed-loop output from one in-house exoskeleton recording. \textbf{(a)}~Multi-modal signals on the shared LSL clock: 64-channel EEG (central-parietal cluster shown, 500~Hz), three-channel surface EMG (2148~Hz), and exoskeleton kinematics (joint torque and position, 100~Hz) around movement onset (dashed line). EEG and EMG are band-pass filtered for display in the band their benchmark decoders use (0.5--10~Hz and 20--200~Hz respectively). \textbf{(b)}~Closed-loop neurofeedback from a study session. The streamed cluster $\alpha$-band-power feature rises during the feedback blocks and relaxes between them, and the app's self-reported end-to-end latency pooled across all live neurofeedback sessions ($p_{50}\approx 13$~ms, $p_{99}\approx 66$~ms) sits below the asynchronous step interval.}
\label{fig:daq-showcase}
\end{figure}

The same harness ran against four datasets. An in-house motor-imagery study (four sessions, single subject, 64 EEG channels at 500~Hz) provides two motor-imagery cue classes, and one replay fits an MI decoder (CSP+LDA). BNCI 2014-001 (the BCI Competition IV-2a dataset~\citep{tangermann2012bci}, with nine subjects, both sessions concatenated, 22 EEG channels at 250~Hz, 4-class MI, and an 8--30~Hz band-pass) is loaded through MOABB~\citep{jayaram2018moabb}. BNCI 2014-009~\citep{riccio2013p300}, a ten-subject P300 oddball dataset (256~Hz, 2-class target/non-target), is loaded the same way and decoded with xDAWN-covariance tangent-space logistic regression. An in-house exoskeleton study contributes a multimodal recording of an exoskeleton-driven motor task, with EEG, EMG, and exoskeleton kinematics captured together (64 EEG channels at 500~Hz). The first three datasets drive the online-learning panels that follow, while the exoskeleton recording drives the acquisition showcase below. Beyond these four datasets, a set of live neurofeedback sessions (five subjects, recorded on the acquisition hardware during $\alpha$-band pretraining) supplies the closed-loop latency reported in Section~\ref{sec:latency}, measured end-to-end through an external feedback application over the LSL contract. The harness reproduces end-to-end with one runner per dataset. Figure~\ref{fig:daq-showcase} shows multi-modal acquisition and the closed-loop neurofeedback path. Rather than establishing new decoding records, these datasets test whether the same runtime can support online learning, offline replay, multimodal acquisition, and closed-loop output across representative BCI paradigms.

\subsection{Online learning across benchmarks}

\begin{figure}[tbp]
\centering
\includegraphics[width=\textwidth]{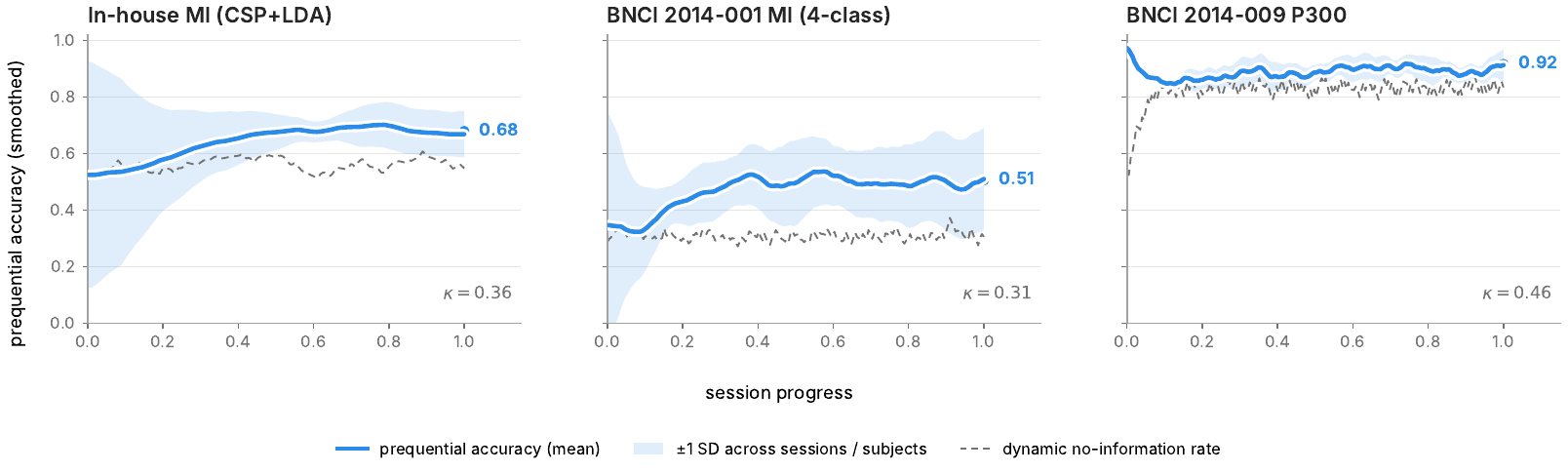}
\caption{Online learning across three benchmark panels: in-house MI (CSP+LDA), BNCI 2014-001 MI, and BNCI 2014-009 P300. Each panel shows the synchronous decoder's prequential accuracy (bold mean with a $\pm1$~SD band across sessions or subjects; 15-sample rolling mean) over normalised session progress, with mean Cohen's $\kappa$ and a dynamic no-information rate (dashed; majority-class proportion under the same $0.95$ forgetting as the accuracy). Series are defined in the figure key.}
\label{fig:bench-online-learning}
\end{figure}

Figure~\ref{fig:bench-online-learning} traces the synchronous decoder's prequential accuracy across the panels. Every value is an online prequential score. Each trial is predicted before its label is seen and then folded into an incrementally retrained decoder under $0.95$ forgetting, so the curves reflect live operation rather than a static fit. The point is that the system produces above-chance decoding while running online, each retrain taking 0.3--0.4~s before the new decoder hot-swaps into the asynchronous mode on its next step.

Across the motor-imagery study's four sessions the MI decoder reaches 0.59--0.82 prequential accuracy ($\kappa$ 0.28--0.46, mean 0.36).

On BNCI 2014-001 (4-class MI, nine subjects of BCI Competition IV-2a), the CSP+LDA pipeline reaches 0.51 mean prequential accuracy and mean $\kappa$ 0.31 (per-subject range 0.08--0.50) against four-class chance of 0.25. The per-subject spread matches the between-subject variability this dataset is known for. On BNCI 2014-009 (P300), the xDAWN-covariance tangent-space pipeline reaches 0.92 mean prequential accuracy and mean $\kappa$ 0.46 (per-subject range 0.14--0.69) across all ten subjects, online-retrained on the same ten-epoch schedule. Raw accuracy is inflated by the 1:5 target/non-target imbalance (no-information rate 0.83), so $\kappa$ is the more honest summary.

\subsection{Real-time performance: latency, cadence, and uninterrupted decoder updates}
\label{sec:latency}

\begin{figure}[tbp]
\centering
\includegraphics[width=\textwidth]{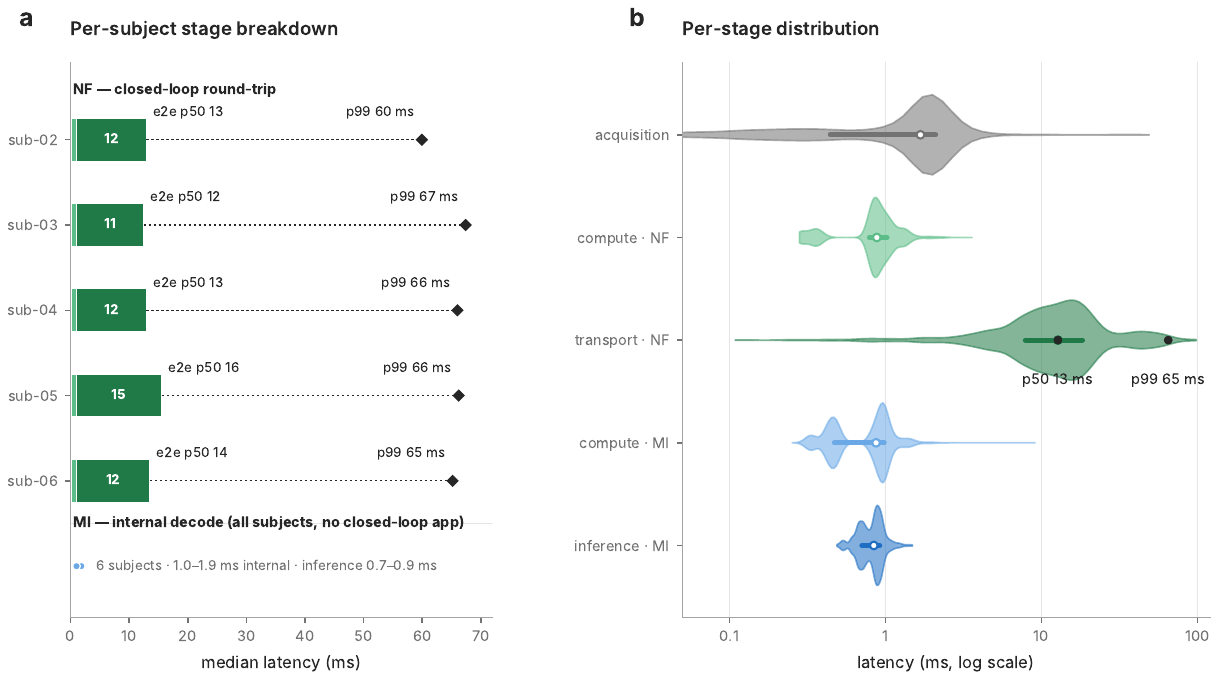}
\caption{Closed-loop latency by stage and subject. \textbf{(a)}~Per-subject median breakdown. Neurofeedback sessions (five subjects, live acquisition hardware) resolve the end-to-end latency into acquisition, backend compute, and transport plus application, with end-to-end $p_{50}$ and $p_{99}$ marked; the transport tail ($p_{99}\approx 60$--67~ms) is consistent across subjects. Motor-imagery sessions ran without an external application and carry only the internal decode stages (acquisition and compute, near 1~ms). \textbf{(b)}~Per-stage distributions pooled across all subjects, log scale. Acquisition and compute stay near or below 1~ms; the latency variability lives in transport.}
\label{fig:latency}
\end{figure}

The asynchronous mode runs on a configurable step grid of 100~ms in the motor-imagery runs and a sample-aligned ${\sim}97.7$~ms in the 256~Hz P300 runs. The wall-clock time of each inference step, recorded in the same telemetry, depends on the decoder. The lightweight CSP+LDA pipelines run at 1.0--1.7~ms at $p_{50}$ and 1.8--2.6~ms at $p_{99}$, while the heavier xDAWN tangent-space P300 pipeline runs at 1.4--2.5~ms at $p_{50}$ and 3.0--3.8~ms at $p_{99}$, with LSL acquisition contributing a further $\sim$0.15~ms at $p_{50}$. Online retraining and hot-swap run on a separate subprocess from the asynchronous loop, so cadence and inference latency are unaffected by training events. These figures are pipeline compute only. The closed-loop latency an external application actually observes, measured end-to-end on the live acquisition hardware and pooled across all neurofeedback sessions, is $p_{50}\approx 13$~ms and $p_{99}\approx 66$~ms (Figure~\ref{fig:daq-showcase}b), dominated by transport and the application's own receive path rather than by inference. The application measures this as the difference between the LSL-synchronised acquisition time of the source EEG sample and its local clock when the derived band-power feature arrives, so the figure spans acquisition, backend compute, and transport rather than inference alone. Figure~\ref{fig:latency} decomposes it per subject and per stage. The internal decode path stays near 1~ms while the latency variability lives in transport, consistent across subjects. This latency stays below the 100~ms asynchronous step and well under the 4~Hz neurofeedback feature update, so end-to-end latency is not a bottleneck for either paradigm.

\FloatBarrier

\section{Discussion}
\label{sec:discussion}

Dendrite addresses a practical biomedical-computing problem in online BCI research, the fragmentation of an online session across separate acquisition, training, and analysis tools. It provides a launchable runtime for acquisition, processing, decoder training, live inference, replay, and data exploration, while keeping the experimental paradigm external through an LSL event contract. This separation is deliberate, and it has costs. Because the paradigm is external, Dendrite includes no stimulus presentation software, so a study must supply its own task application, and the initial setup is heavier than on a platform that bundles one. The same decoupling extends to the pipeline, where each stream is decoded independently and predictions are combined only at the end. This allows any LSL stream to join as a separate mode but prevents a decoder from learning features jointly across modalities.

The evaluation tested end-to-end operation. Across the in-house motor-imagery, BNCI 2014-001, and BNCI 2014-009 P300 benchmarks, online-trained decoders stayed above chance on all three. Training inside the live pipeline yields sensible decoders instead of disrupting them. The policy refits the decoder from scratch every ten epochs and hot-swaps the result into the running asynchronous mode, and more sophisticated online-learning methods, from incremental and adaptive updates to cross-session transfer, can be built on top. The acquisition showcase captured EEG, EMG, and exoskeleton kinematics together on the shared LSL clock, each at its native rate (500, 2148, and 100~Hz), with per-stream ring buffers keeping them aligned (Figure~\ref{fig:daq-showcase}a). The closed-loop neurofeedback path, driving an external application over LSL with no platform-specific adapter, exercised the event-and-prediction contract end-to-end (Figure~\ref{fig:daq-showcase}b). Throughout, the runtime met its real-time obligations. Per-step inference stayed between 1 and 4~ms, well inside the asynchronous step interval, and the closed-loop latency an external application observed was $p_{50}\approx 13$~ms, with online retraining running off the inference path so a training event never perturbed cadence (Section~\ref{sec:latency}).

Extending the runtime means editing its source. A new decoder registers in the model registry as a local change; a new processing mode or output protocol is added by editing the layer that owns it. Dendrite exposes no plugin API, so a new capability is a modification in Python, the language the runtime is already written in, with the whole source open to change. The cost is that such extensions cannot be shipped as standalone plugin packages; they live in a fork or upstream. This runs in two directions. Inside the codebase, a researcher can assemble a custom paradigm or drop in an experimental decoder to probe a single stage of the pipeline in isolation. Because modes are composed declaratively rather than wired into a fixed pipeline, many decoder and channel configurations can run concurrently against the same acquisition stream, so experimental capacity is set by configuration rather than by new code. Outside the codebase, the FastAPI REST layer that drives the web interface is equally reachable by any programmatic client, so nothing about the system is exclusive to the browser UI. The same actions an operator performs there (discovering streams, configuring modes, launching the pipeline, reading back predictions) are available to a script or an autonomous agent. This makes the research loop itself automatable. A client can sweep configurations, launch replay runs, and collect results with no human in the loop, and because every run reproduces from its saved configuration and every trained decoder stores the configuration and training run that produced it together with the identifiers of its source recordings, the outcomes of such a sweep stay traceable to what produced them.

The present evaluation has several limitations. The online-learning policy is intentionally simple and was used to exercise the runtime rather than to advance adaptive BCI algorithms. The in-house datasets are modest in size, and the closed-loop neurofeedback latency was measured with one external feedback application rather than across many independent clients. Finally, multimodal fusion currently occurs downstream of independent per-modality decoders, and Dendrite does not yet train joint multimodal models inside the runtime.

\section*{Conclusions}

Dendrite is an open-source Python application for reproducible online BCI research. It integrates multimodal acquisition, real-time processing, online decoder training, live inference, offline replay, browser-based monitoring, and traceable data storage in a single runtime while keeping task paradigms external through an LSL event contract. Across public and in-house datasets, Dendrite supported online decoder updates, concurrent processing modes, and closed-loop output with millisecond-scale inference latency. The application is intended to reduce the software burden of developing and evaluating online BCI paradigms while preserving the flexibility of the scientific Python ecosystem.

\section*{Code availability}

Dendrite is released as open-source software under the GPL-3.0 license at \url{https://github.com/dendrite-bci/dendrite}. This paper describes release v0.10.

\bibliographystyle{plainnat}
\bibliography{references}

\end{document}